# Non-collapsibility and Built-in Selection Bias of Hazard Ratio in Randomized Controlled Trials

Helen Bian[1], Menglan Pang[2], Guanbo Wang[3,4], Zihang Lu[5]

[1]Department of Epidemiology, Biostatistics and Occupational Health, McGill University, Montreal, Canada

[2]Biogen, Cambridge, USA

[3]CAUSALab, Harvard T.H. Chan School of Public Health, Cambridge, USA

[4]Department of Epidemiology, Harvard T.H. Chan School of Public Health, Cambridge, USA

[5]Department of Public Health Sciences & Department of Mathematics and Statistics, Queen's University, Kingston, Canada

Correspondence to: Helen Bian

Mailing Address: 2001 McGill College, Suite 1200, Montreal, QC, Canada H3A 1G1

E-mail: he.bian@mail.mcgil.ca

Phone Number: (+1)-416-893-3787

**Abstract**

**Background** The hazard ratio of the Cox proportional hazards model is widely used in randomized controlled trials to assess treatment effects. However, two properties of the hazard ratio including the non-collapsibility and built-in selection bias need to be further investigated.

**Methods** We conduct simulations to differentiate the non-collapsibility effect and built-in selection bias from the difference between the marginal and the conditional hazard ratio. Meanwhile, we explore the performance of the Cox model with inverse probability of treatment weighting for covariate adjustment when estimating the marginal hazard ratio. The built-in selection bias is further assessed in the period-specific hazard ratio.

**Results** The conditional hazard ratio is a biased estimate of the marginal effect due to the non-collapsibility property. In contrast, the hazard ratio estimated from the inverse probability of treatment weighting Cox model provides an unbiased estimate of the true marginal hazard ratio. The built-in selection bias only manifests in the period-specific hazard ratios even when the proportional hazards assumption is satisfied. The Cox model with inverse probability of treatment weighting can be used to account for confounding bias and provide an unbiased effect under the randomized controlled trials setting when the parameter of interest is the marginal effect.

**Conclusions** We propose that the period-specific hazard ratios should always be avoided due to the profound effects of built-in selection bias.

**Keywords** Non-collapsibility, Built-in selection bias, Cox proportional hazards model, Inverse probability of treatment weighting, Randomized controlled trials.

**Background**

The data with time-to-event outcomes collected from randomized controlled trials (RCTs) have been widely used to assess the treatment effect of new interventions on the target population in many therapeutic areas, such as oncology, cardiology, and hematology, etc. To analyze this type of data, the Cox proportional hazards (PH) model [1] has become a common approach and has been implemented in many standard statistical software. In the standard Cox PH model, the hazard ratio (HR) which represents the coefficient of the treatment variable, is assumed to be constant over time and is often reported as the treatment effect in an RCT. In RCTs, the marginal HR is usually chosen as one of the target parameters for measuring the efficacy of treatment effects at the population level providing crucial support for drug approval.

It is well known that the HR is not generally collapsible [2,3]. The non-collapsibility is a property of a coefficient in a statistical model. When focusing on the coefficient of the treatment and collapsing over covariate(s), the conditional effect does not equal the marginal effect even in the absence of confounding and effect modification [4-6]. This is because the relationship between the covariates and hazard is assumed non-linear in the Cox PH model [2,7]. Gail et al. showed that the marginal HR is always shifted towards the null compared to the conditional HR when covariates are omitted, and the magnitude depends on the strength of the treatment effect, the effects of covariates on the outcome, and the censoring events [2]. Moreover, the non-collapsibility effect may occur in the absence of confounders, and confounding bias can occur without non-collapsibility [5,6,8-10]. Because of this property, many researchers have pointed out the issue of quantifying the confounding effects of the factors by comparing the estimates of regression coefficients with or without adjustment for the factors [5,11,12]. More recently, the non-collapsible property of the HR and odds ratio has been discussed in the FDA guidelines [13], which emphasizes the importance of differentiating subgroup-specific conditional treatment effects from unconditional treatment effects even when all subgroup treatment effects are identical for the RCT. Meanwhile, it is important to distinguish the non-collapsibility effects and confounding effects, as the former can persist despite randomization and large sample size [14]. On the other hand, Hernán has raised critical concerns about the HR estimated from the Cox PH model based on two main reasons [15]. First, when the PH assumption is violated, the single HR is an average effect over the duration of the study's follow-up time, without considering the distribution of events. Second, period-specific HRs, estimated based on subjects who are free of events at the beginning of specific periods, are proposed to address the time-dependent hazard ratio, yet previous research has shown that period-specific HRs are subject to built-in selection bias [15,16,17]. In the RCT, all the baseline characteristics, such as harmful and protective factors, including both measured and unmeasured, are expected to be equally distributed among those groups through randomization. However, the balance of these covariates can only be obtained at baseline and will be easily broken at post-randomization period among survivors, because the speed of the outcome occurrence differs between treatment groups, especially when the treatment effects are unignorable or the outcome rate is not rare [15,17]. This phenomenon is also referred to as the differential depletion of susceptibles between treatment groups [15]. As a result, risk factors, if not being appropriately controlled for, may become confounders over time, which induces the built-in selection bias in period-specific HR. On the other hand, in the RCT, only a limited number of clinical prognostic factors are being measured and adjusted. Because of the differential depletion of the susceptibles between the treatment groups especially with respect to unmeasured covariates, it was pointed out in the literature that the contribution to the partial likelihood in the Cox PH model is not completely based on the randomized comparison [16]. Due to the presence of the selection bias, there have been increased discussions regarding the causal interpretation of the marginal HR for the RCT [16-18]. Particularly, the Cox hazard ratio is not causally interpretable as a hazard ratio unless there is no treatment effect or an untestable and unrealistic assumption holds.[18] Nevertheless, when the PH assumption is reasonably satisfied over the fixed follow-up time, how much the single estimated marginal HR will suffer from the selection bias has not been well understood.

The objective of this paper is to assess the non-collapsibility effect and the potential built-in selection bias of the HR simultaneously in the Cox PH model under the PH assumption in the RCT setting. The collapsibility can be defined based on different approaches of marginalizing or standardizing over the covariate. However, in this paper, we focus on estimating the HR of the Cox PH model with and without the adjustment for one risk factor.

**Methods**

*Evaluation of non-collapsibility and built-in selection bias*

Since both the non-collapsibility effects and built-in selection bias may affect the comparison of the estimated HR with or without adjusting for covariates, we propose to evaluate these two effects simultaneously.

First, the hazard function of the Cox PH model is given by:

$$h(t|z,l) = h_0(t) \times \exp(\beta_0 z + \beta_1' l')$$

where $z$ is the treatment, $l'$ is the vector of covariates. The conditional HR (denoted by $HR_{Adj}$) estimated by the Cox model with the covariates adjusted in the regression as $\exp(\hat{\beta}_0)$ reveals an conditional estimate of the treatment effect, which is biased from the marginal effect due to non-collapsibility [2], even with all the confounders adjusted in the model. On the other hand, omitting any covariate in the regression provides the crude HR (denoted by $HR_{Unadj}$) which might be subject to the built-in selection bias in the RCT [15-17]. In order to distinguishing the non-collapsibility effect and the built-in selection bias, the true marginal HR (denoted by $HR_M$) is introduced. The true marginal HR has the causal interpretation of treatment effects by comparing the hazard function of failure time between the same treated and untreated individuals.

Therefore, if the true marginal HR is known, we can quantify the non-collapsibility effect of the HR by comparing the $HR_{Adj}$ to $HR_M$ in the log scale and measure the built-in selection bias of the HR by comparing the $HR_M$ to $HR_{Unadj}$ in the log scale as the decomposition formula (1) [5,19].

$$\log(HR_{Adj}) - \log(HR_{Unadj}) = \underbrace{[\log(HR_{Adj}) - \log(HR_M)]}_{\text{Non-collapsibility effects}} + \underbrace{[\log(HR_M) - \log(HR_{Unadj})]}_{\text{Built-in selection bias}} \quad (1)$$

The framework was initially proposed to distinguish the magnitude of confounding from the size of the non-collapsibility effects [19]. Here, we utilize it to quantify the built-in selection bias from the Cox model under the RCT setting. Since it is reasonable to assume the confounding bias is excluded through the randomization, the total discrepancy between the $HR_{adj}$ and $HR_{unadj}$ in the log scale can be consequently decomposed into two components, namely the non-collapsibility effects and a built-in selection bias.

*Simulation*

Analytical derivation of the non-collapsibility effect as well as a built-in selection bias of the HR can be relatively complicated; therefore, a simulation study was used to investigate both properties of the HR when a harmful or protective factor was omitted.

To evaluate the non-collapsibility effect and the built-in selection bias of the HR, we considered a simulation study, in which $N = 10,000$ subjects were randomized with 1:1 ratio to either the control ($Z = 0$) or treated group ($Z = 1$). Each subject $i$ had the same probability of receiving either placebo or treatment as the treatment variable $Z_i$ was generated from a Bernoulli distribution ($P[Z_i(1)] = 0.5$). The maximum follow-up time was 3 years, and the outcome of interest was time to death. Denote $T_i$ as the observed survival time of subject $i$. We considered only one baseline harmful or protective factor in the study, denoted as covariate $L$. We focused on the scenario in which the HR was constant over time satisfying the PH assumption. The baseline factor $L_i$ was randomly generated for each subject from a normal distribution $N(5, 2)$. Survival times $T_i$ was generated from a PH model, by inversing the cumulative baseline hazard function [20], i.e.,

$$T_i = H_0^{-1}[-\log(U) \times \exp(-\beta_1 Z_i - \beta_2 L_i)] \quad (4)$$

where U was a random variable that follows a uniform distribution $U(0,1)$.

We assumed that the baseline hazard function follows the Weibull distribution with scale parameters λ=0.0001 and shape ν=2, then the survival time can be expressed as follows:

$$T_i = \left[\frac{-\log(U)}{\lambda \times \exp(\beta_1 Z_i + \beta_2 L_i)}\right]^{\frac{1}{\nu}}. \quad (5)$$

Censoring time $C_i$ was independently generated from an exponential distribution with rate $\lambda = 0.005$. Let $T_i^* = \min\{T_i, C_i\}$ denote the observed time which was the minimum of the event time and the censoring time. The effect of the treatment ($HR_E$, i.e., $\exp(\beta_0)$) was assigned to be one of the values in the set {0.41, 0.61, 1, 1.65, 2.46} corresponding to the $log\ (HR_E)$ {-0.9, -0.5, 0, 0.5, 0.9} to reflect the various treatment effects. The harmful or protective factor ($HR_L$) was assigned to one of the values in the set {0.67, 0.82, 1, 1.22, 1.49} corresponding to the $log\ (HR_L)$ {-0.4, -0.2, 0, 0.2, 0.4}. In total, there were 25 different parameter settings. All the simulations were conducted by using R software version 4.1.3 [21].

The above data generating mechanism ensures that the true conditional HR is the value assigned in the parameter setting, however the true marginal HR is still unknown. We therefore used separate simulations for each parameter setting to determine the corresponding true marginal effect that was induced by the specified conditional effect. We first simulated *N*=10,000 subjects following the data generation procedure described above, and subsequently generated another *N* subjects as the duplicates of the first *N* subjects. These duplicates inherited the values of the baseline factor from the original subjects, however, had completely opposite treatment assignments. Thus, the true marginal HR also known as the average treatment effect of moving an entire population from untreated to treated can be attained [22]. For each simulation, we repeated this process 100 times and obtained the mean of the effects estimated from the Cox model. This mean estimate was then considered as the true value of the marginal HR, because every subject contributed twice to the study with and without receiving the treatment, and no confounding was present in the simulated sample.

*Propensity score methods*

Another way to estimate the true marginal effect is to use the propensity score method, while adjusting for confounding factors via the inverse probability of treatment weighting (IPTW) [23,24]

The true marginal HR is not easy to derive, when the data generating mechanism relies on the distribution of event time conditional on covariates. However, propensity score (PS) methods can be used to estimate the true marginal HR, while adjusting for confounding factors via the inverse probability of treatment weighting (IPTW) [23,24].

It is commonly used for the observational studies where the balance of covariates is assumed to be broken compared to that in RCTs. Here we used this method to estimate the true marginal effect by adjusting the prognostic factors which are known relevant to the outcome and may bring the selection bias if not appropriately adjusted. The PS is defined as the probability of receiving the treatment given covariates. It is considered a summary score for confounding adjustment and is particularly useful when there are many covariates to be considered in a study [25]. The PS is classically estimated with a logistic regression conditional on covariates. In the context of our simulation study, the covariate was the baseline factor $L$. Explicitly, the estimated propensity score can be expressed as: $\widehat{PS} = \frac{exp(\hat{\alpha}_0 + \hat{\alpha}_1 L)}{1 + exp(\hat{\alpha}_0 + \hat{\alpha}_1 L)}$.

The marginal HR($e^\beta$) can be estimated from the IPTW Cox proportional hazards model via maximizing a weighted partial likelihood:

$$L(\beta) = \prod_{i=1}^{N} \left[ \frac{e^{(\hat{\beta} Z_i)}}{\sum_{j \in R(t)} \widehat{w}_j e^{(\hat{\beta} Z_j)}} \right]^{\widehat{w}_i} \quad (2)$$

where $R$ is the risk set denoting the set of individuals $j$ who are at risk for failure at time $t$, and the estimated weight $\widehat{w}_i$ is calculated from the propensity score. For the treated subjects, the weights are estimated by the inverse of the propensity score, and for the untreated subjects the weights are estimated by the inverse of one minus the propensity score [23].

$$w_i = \frac{Z_i}{p(L_i)} + \frac{(1 - Z_i)}{1 - p(L_i)} \quad (3)$$

where $Z_i$ indicates whether the treatment is received or not, and $p(L_i)$ is the calculated PS for subject $i$.

*Analysis of the Simulated Data*

The simulation was repeated 500 times. Each simulated sample was analyzed using three different models to estimate the HR. The conditional HR and the crude HR were estimated from the Cox model with and without adjusting for $L$ in the regression, providing the $HR_{Adj}$ and $HR_{Unadj}$. Moreover, the marginal HR was estimated through the duplicate samples providing $HR_M$ and from the IPTW Cox model with the weights estimated by the propensity score, providing $HR_{IPWT}$ [23,26,27]. These two pairs of the estimated marginal HR were subsequently compared to check the performance of the $HR_{IPWT}$ in the presence of the non-collapsibility effect.

In addition, the true marginal $HR_M$ estimated through the duplicated samples was used to disassemble the non-collapsibility effects and built-in selection bias from the mean difference of $HR_{Adj}$ and $HR_{Unadj}$ using the decomposition formula (1). The 95% confidence intervals of both effects were also calculated based on the Monte-Carlo standard deviation of the estimates across 500 simulations ($mean \pm sd \times 1.96$).

In addition, the period-specific HR was estimated from an unadjusted Cox model, which was continuously performed for each separate 50-day interval for harmful factor ($HR_L > 0$) and 100-day interval for protective factor respectively. The choice of intervals was influenced by the more rapid depletion of susceptibles exposed to a harmful factor compared to those exposed to a protective factor. The mean value of the factor by the end of each interval was also computed.

**Results**

*Non-collapsibility Effect*

Table 1 presents the estimated the non-collapsibility effect and a built-in selection bias of the HR and their 95% confidence intervals. The estimated conditional HR as $\widehat{HR}_{Adj}$, the estimated crude HR as $\widehat{HR}_{Unadj}$, and the estimated marginal effect as $\widehat{HR}_M$ based on the duplicates were also presented. The results confirmed that the $\widehat{HR}_{Unadj}$ always shifts toward the null compared to the $\widehat{HR}_{Adj}$. This also applied when comparing the $\widehat{HR}_{Adj}$ to $\widehat{HR}_M$, as the $\widehat{HR}_M$ was always closer to 0 compared to the $\widehat{HR}_{Adj}$, i.e., the non-collapsibility effect was positive for a harmful treatment and negative for a protective treatment. There was no non-collapsibility effect whenever either the treatment effect or the factor effect was null. However, it increased with increasing of either one of the effects while the other was fixed. For example, the non-collapsibility effect was -0.001 (-0.082, 0.085) when $HR_E$=2.46 and $HR_L$=1, which was not significantly different from 0 whereas it changed to be significantly different from 0 as 0.227 (0.175, 0.280), when $HR_L$ increased to 1.49.

*Built-in Selection Bias in the Crude HR*

By comparing the $\widehat{HR}_{Unadj}$ to $\widehat{HR}_M$, the built-in selection bias was close to 0 in all the parameter settings with 95% confidence intervals always included the null as results shown in Table 1. This indicated that the built-in selection bias was negligible when the HR was estimated using the entire follow-up data. Meanwhile, the marginal HR estimated from the IPTW Cox model as $\widehat{HR}_{IPTW}$ was nearly close to the $\widehat{HR}_M$ under all the conditions. This confirmed that using the IPTW Cox model to adjust for the covariate(s) can provide unbiased marginal effect for time-to-event data, which would not suffer from the non-collapsibility effect of the HR.

*Built-in Selection Bias in the Period-Specific HR*

Table 2 illustrated the built-in selection bias by examining changes in the period-specific HRs estimated at each 50-day interval for a harmful factor. The estimated period-specific HRs deviated from the $\widehat{HR}_M$ over time and moved toward null during the follow-up period. Moreover, the magnitude of the factor effect influenced the speed of this convergence toward null. Notably, in extreme cases, the direction of the period-specific treatment effect was reversed in later follow-up intervals. For instance, the estimated period-specific treatment effect became protective in the 351-400 day interval, even when the treatment effect was harmful ($\log(HR_E) = 0.5 \text{ and } 0.9$, and $\log(HR_L) = 0.2 \text{ and } 0.4$). Similar patterns were observed, as shown in Table 3.

To assess the imbalance of the factor between the treated and control groups, Figure 1 displayed the mean values of the harmful factor ($HR_L = 1.49$) at the end of each interval among subjects who did not have the outcome event. The red line in the middle represented the mean values of the harmful factor in the control group surrounded by the other four lines represent the mean values of the harmful factor in treated groups with varying strengths of the treatment effects: $HR_E = 0.41, HR_E = 0.61, HR_E = 1.65$ and $HR_E = 2.46$ respectively. All the values started at 5 at the beginning of follow-up, then decreased toward zero over time. This was because the susceptible subjects with higher values of the harmful factor were more likely to be excluded earlier from the risk sets within those two groups receiving harmful treatments, and as a result, the mean values of the risk factor decreased faster in these two groups.

Similarly, Figure 2 displayed the scenarios under the protective factor with $HR_L = 0.67$. All the lines started at 5 then gradually increased over time, which indicated the susceptible subjects with lower values of the protective factor were removed faster during the follow-up. Moreover, the speed of the removal of the susceptibles was protected by protective treatment effects.

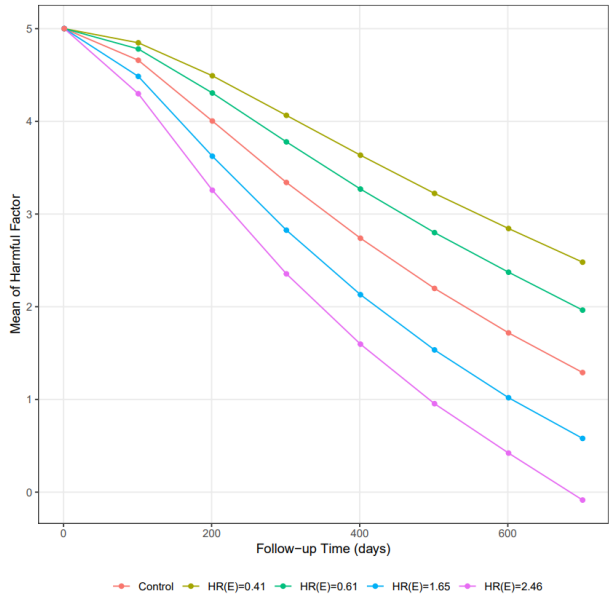

*Figure 1 The mean values of the harmful factor (HR_L=1.49) for the treated and control groups over follow-up.*

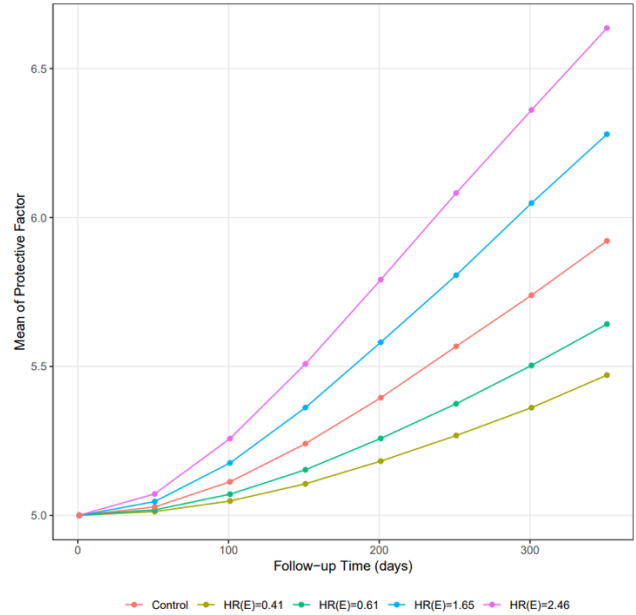

*Figure 2 The mean values of the protective factor (HR_L=0.67) for the treated and control groups over follow-up.*

**Conclusions**

The HR estimated from the Cox model has been widely used in RCTs to analyze time-to-event data. However, care should be taken when interpreting it due to non-collapsibility[2,5] and built-in selection bias[15-17].

In this paper, we have conducted simulations to differentiate the non-collapsibility effects and built-in selection bias from the difference between the marginal HR and the conditional HR for the Cox PH model under the RCT setting. When neither treatment effects nor the factor effects were not null, non-collapsibility effects were not ignorable, and it pulled the conditional HR farther away from the null compared to the marginal HR[2]. Therefore, if the target parameter is the marginal effect, which reflects the average treatment effects at the population level, the conditional HR is a biased estimate of the marginal effect due to the non-collapsibility property. However, the unadjusted Cox model offers unbiased estimates of the marginal effect. Thus, when the parameter of interest is the marginal treatment effect, the unadjusted Cox model should be used and adjusted model should be avoided.

In the meantime, we have shown that the HR estimated from the IPTW Cox model can provide an unbiased estimate of the true marginal HR, which did not suffer from the non-collapsibility effect. This finding is particularly relevant in certain RCTs, where concerns arise about the blend in randomization factor and prolonged effects of prognostic factors. To address these concerns, the randomization is often stratified by baseline covariates, which a covariate adjustment model should generally include strata variables [13]. With noticing the non-collapsibility effect of the HR, the IPTW can be considered which provides reliable estimate of average treatment effects. It was also shown in our simulations that although the built-in selection bias has been recognized [15,17], it is only subject to the period-specific HRs but not to the unadjusted HR estimated for a fixed duration of follow-up. According to our simulation results, given the fixed observation time, the HR less suffered from a built-in selection bias if the PH assumption can be held. On the other hand, the variation of the period-specific HRs induced by the imbalance of the risk factor between the control and treated groups over time has been identified. Therefore, we propose that the period-specific HRs should be always avoided in survival analysis, irrespective of whether the PH assumption is satisfied or not, due to the profound effects of the built-in selection bias.

In summary, the HR of the Cox model should be carefully interpreted concerning its non-collapsibility feature and a built-in selection bias [16,17]. In RCTs, the marginal HR is a valid interpretation, providing the causal estimate of the average treatment effect at the population-level for a fixed follow-up time. The difference between the conditional and marginal HR should not be wrongly interpreted as confounding effects. In addition, period-specific HRs could be severely affected by the built-in selection bias, which could lead to inaccurate and even opposite conclusions.

**Tables**

**Table 1.** The estimates and 95% confidence intervals of the non-collapsibility effect and built-in selection bias with respect to the continuous time-fixed factor.

| Treatment effect | Factor effect | Estimated marginal effect | Estimated conditional effect | Estimated crude effect | Estimated effect of IPTW Cox model | Non-collapsibility effect | Built-in selection bias |
|---|---|---|---|---|---|---|---|
| $log(HR_E)$ | $log(HR_L)$ | $log(\widehat{HR}_M)$ | $log(\widehat{HR}_{adj})$ (95% CI) | $log(\widehat{HR}_{unadj})$ (95% CI) | $log(\widehat{HR}_{IPTW})$ (95% CI) | $log(\widehat{HR}_{adj}) - log(\widehat{HR}_M)$ (95% CI) | $log(\widehat{HR}_M) - log(\widehat{HR}_{unadj})$ (95% CI) |
| -0.9 | -0.4 | -0.823 | -0.901 (-1.053, -0.75) | -0.824 (-0.972, -0.676) | -0.823 (-0.967, -0.679) | -0.079 (-0.23, 0.073) | 0.001 (-0.147, 0.149) |
|  | -0.2 | -0.865 | -0.897 (-1.009, -0.784) | -0.861 (-0.973, -0.748) | -0.860 (-0.971, -0.750) | -0.032 (-0.145, 0.08) | -0.004 (-0.117, 0.109) |
|  | 0 | -0.900 | -0.899 (-0.982, -0.815) | -0.899 (-0.982, -0.815) | -0.899 (-0.982, -0.815) | 0.001 (-0.082, 0.085) | -0.002 (-0.085, 0.082) |
|  | 0.2 | -0.833 | -0.899 (-0.963, -0.834) | -0.832 (-0.897, -0.767) | -0.832 (-0.895, -0.769) | -0.066 (-0.131, -0.001) | -0.001 (-0.066, 0.064) |
|  | 0.4 | -0.688 | -0.899 (-0.955, -0.844) | -0.687 (-0.742, -0.632) | -0.687 (-0.736, -0.638) | -0.212 (-0.267, -0.156) | -0.001 (-0.056, 0.054) |
| -0.5 | -0.4 | -0.453 | -0.502 (-0.636, -0.367) | -0.454 (-0.588, -0.321) | -0.454 (-0.583, -0.325) | -0.048 (-0.183, 0.087) | 0.001 (-0.028, 0.028) |
|  | -0.2 | -0.479 | -0.498 (-0.6, -0.397) | -0.476 (-0.578, -0.374) | -0.476 (-0.576, -0.376) | -0.020 (-0.122, 0.082) | -0.002 (-0.105, 0.100) |
|  | 0 | -0.500 | -0.499 (-0.576, -0.422) | -0.499 (-0.576, -0.422) | -0.499 (-0.576, -0.422) | 0.001 (-0.076, 0.078) | -0.001 (-0.078, 0.076) |
|  | 0.2 | -0.461 | -0.499 (-0.559, -0.439) | -0.460 (-0.521, -0.399) | -0.460 (-0.519, -0.401) | -0.039 (-0.099, 0.022) | -0.001 (-0.062, 0.06) |
|  | 0.4 | -0.379 | -0.500 (-0.552, -0.448) | -0.379 (-0.43, -0.327) | -0.379 (-0.425, -0.333) | -0.121 (-0.173, -0.069) | 0.000 (-0.052, 0.051) |
| 0 | -0.4 | 0.000 | -0.001 (-0.125, 0.123) | -0.001 (-0.125, 0.123) | -0.001 (-0.119, 0.118) | -0.001 (-0.125, 0.123) | 0.001 (-0.028, 0.027) |
|  | -0.2 | 0.000 | 0.002 (-0.089, 0.092) | 0.002 (-0.088, 0.092) | 0.002 (-0.087, 0.091) | 0.002 (-0.089, 0.092) | -0.002 (-0.015, 0.015) |

| log($HR_E$) | log($HR_L$) | log($\widehat{HR}_M$) | Est 1 | Est 2 | Est 3 | Est 4 | Est 5 |
|---|---|---|---|---|---|---|---|
|  | 0 | 0.000 | 0.000 | 0.000 | 0.000 | 0.000 | 0.000 |
|  |  |  | (-0.071, 0.071) | (-0.071, 0.071) | (-0.071, 0.071) | (-0.071, 0.071) | (-0.001, 0.001) |
|  | 0.2 | 0.000 | 0.000 | 0.000 | 0.000 | 0.000 | 0.000 |
|  |  |  | (-0.057, 0.056) | (-0.057, 0.057) | (-0.055, 0.055) | (-0.057, 0.056) | (-0.014, 0.014) |
|  | 0.4 | 0.000 | 0.000 | 0.000 | 0.000 | 0.000 | 0.000 |
|  |  |  | (-0.050, 0.050) | (-0.049, 0.05) | (-0.044, 0.044) | (-0.050, 0.050) | (-0.023, 0.023) |
| 0.5 | -0.4 | 0.442 | 0.499 | 0.441 | 0.441 | 0.057 | 0.001 |
|  |  |  | (0.384, 0.614) | (0.327, 0.554) | (0.332, 0.550) | (-0.058, 0.172) | (-0.028, 0.027) |
|  | -0.2 | 0.474 | 0.501 | 0.475 | 0.474 | 0.027 | -0.001 |
|  |  |  | (0.381, 0.621) | (0.355, 0.594) | (0.358, 0.590) | (-0.094, 0.147) | (-0.02, 0.021) |
|  | 0 | 0.500 | 0.501 | 0.501 | 0.501 | 0.001 | 0.000 |
|  |  |  | (0.405, 0.597) | (0.405, 0.597) | (0.405, 0.597) | (-0.096, 0.097) | (-0.001, 0.001) |
|  | 0.2 | 0.457 | 0.500 | 0.457 | 0.457 | 0.043 | 0.000 |
|  |  |  | (0.443, 0.556) | (0.401, 0.513) | (0.402, 0.511) | (-0.014, 0.099) | (-0.014, 0.014) |
|  | 0.4 | 0.374 | 0.499 | 0.374 | 0.374 | 0.125 | 0.000 |
|  |  |  | (0.449, 0.550) | (0.325, 0.423) | (0.329, 0.418) | (0.074, 0.176) | (-0.023, 0.023) |
| 0.9 | -0.4 | 0.786 | 0.899 | 0.785 | 0.786 | 0.113 | 0.001 |
|  |  |  | (0.789, 1.009) | (0.677, 0.894) | (0.682, 0.889) | (0.002, 0.223) | (-0.027, 0.027) |
|  | -0.2 | 0.849 | 0.900 | 0.849 | 0.849 | 0.051 | -0.001 |
|  |  |  | (0.817, 0.983) | (0.767, 0.932) | (0.768, 0.931) | (-0.032, 0.135) | (-0.015, 0.014) |
|  | 0 | 0.900 | 0.899 | 0.899 | 0.899 | -0.001 | 0.001 |
|  |  |  | (-0.982, -0.815) | (-0.982, -0.815) | (-0.982, -0.815) | (-0.082, 0.085) | (-0.001, 0.001) |
|  | 0.2 | 0.823 | 0.900 | 0.822 | 0.822 | 0.077 | 0.000 |
|  |  |  | (0.843, 0.956) | (0.766, 0.879) | (0.767, 0.877) | (0.02, 0.134) | (-0.014, 0.014) |
|  | 0.4 | 0.672 | 0.899 | 0.672 | 0.671 | 0.227 | 0.000 |
|  |  |  | (0.847, 0.952) | (0.622, 0.722) | (0.626, 0.717) | (0.175, 0.28) | (-0.023, 0.024) |

**Table 2.** The estimates of period-specific HR of 50-days interval with harmful factor ($log\ (HR_L) > 0$).

| Treatment effect | Risk factor effect | True marginal effect | Estimated period-specific effect (day) | | | | | | | |
|---|---|---|---|---|---|---|---|---|---|---|
|  |  |  | $log(\widehat{HR}_{unadj})$ | | | | | | | |
|  |  |  | (95% CI) | | | | | | | |
| $log\ (HR_E)$ | $log\ (HR_L)$ | $log(\widehat{HR}_M)$ | 1 - 50 | 51 - 100 | 101 - 150 | 151 - 200 | 201 - 250 | 251 - 300 | 301 - 350 | 351 - 400 |
| -0.9 | 0.2 | -0.833 | -0.9 | -0.87 | -0.84 | -0.81 | -0.77 | -0.73 | -0.69 | -0.64 |
|  |  |  | (-1.102, -0.692) | (-1.017, -0.732) | (-0.975, -0.713) | (-0.961, -0.655) | (-0.951, -0.581) | (-0.983, -0.484) | (-1.015, -0.357) | (-1.140, -0.144) |
|  | 0.4 | -0.688 | -0.84 | -0.71 | -0.59 | -0.52 | -0.45 | -0.42 | -0.38 | -0.24 |
|  |  |  | (-0.955, -0.724) | (-0.807, -0.616) | (-0.712, -0.477) | (-0.678, -0.355) | (-0.689, -0.22) | (-0.797, -0.043) | (-0.991, 0.24) | (-2.715, 2.229) |
| -0.5 | 0.2 | -0.461 | -0.498 | -0.482 | -0.461 | -0.443 | -0.416 | -0.394 | -0.365 | -0.334 |
|  |  |  | (-0.681, -0.314) | (-0.606, -0.357) | (-0.584, -0.338) | (-0.588, -0.297) | (-0.587, -0.245) | (-0.633, -0.156) | (-0.700, -0.030) | (-0.857, 0.188) |
|  | 0.4 | -0.379 | -0.461 | -0.38 | -0.318 | -0.273 | -0.236 | -0.225 | -0.204 | -0.09 |
|  |  |  | (-0.564, -0.358) | (-0.472, -0.289) | (-0.426, -0.21) | (-0.431, -0.115) | (-0.492, 0.019) | (-0.599, 0.148) | (-0.851, 0.444) | (-2.67, 2.49) |
| 0 | 0.2 | 0 | -0.002 | 0.006 | -0.001 | -0.001 | 0.015 | -0.011 | -0.008 | 0.049 |
|  |  |  | (-0.158, 0.155) | (-0.107, 0.112) | (-0.113, 0.118) | (-0.138, 0.127) | (-0.180, 0.176) | (-0.266, 0.252) | (-0.388, 0.395) | (-0.608, 0.615) |
|  | 0.4 | 0.000 | -0.002 | 0.006 | -0.001 | -0.001 | 0.015 | -0.011 | -0.008 | 0.049 |
|  |  |  | (-0.089, 0.091) | (-0.085, 0.087) | (-0.106, 0.104) | (-0.168, 0.17) | (-0.269, 0.273) | (-0.471, 0.426) | (-0.794, 0.763) | (-3.343, 3.444) |
| 0.5 | 0.2 | 0.457 | 0.492 | 0.475 | 0.444 | 0.407 | 0.375 | 0.327 | 0.311 | 0.178 |
|  |  |  | (0.352, 0.632) | (0.371, 0.580) | (0.341, 0.546) | (0.263, 0.551) | (0.178, 0.573) | (0.022, 0.630) | (-0.200, 0.822) | (-2.730, 3.086) |
|  | 0.4 | 0.374 | 0.443 | 0.344 | 0.277 | 0.232 | 0.198 | 0.174 | 0.149 | -2.218 |
|  |  |  | (0.361, 0.525) | (0.263, 0.425) | (0.164, 0.390) | (0.041, 0.424) | (-0.146, 0.542) | (-0.400, 0.747) | (-0.968, 1.265) | (-15.536, 11.1) |
| 0.9 | 0.2 | 0.823 | 0.883 | 0.842 | 0.775 | 0.703 | 0.638 | 0.556 | 0.496 | -1.76 |
| | | | (0.755, 1.011) | (0.743, 0.941) | (0.678, 0.873) | (0.560, 0.845) | (0.414, 0.863) | (0.132, 0.981) | (-1.149, 2.142) | (-13.512, 10.01) |
| | 0.4 | 0.672 | 0.779 | 0.589 | 0.468 | 0.388 | 0.361 | 0.237 | -1.788 | -6.11 |
| | | | (0.702, 0.856) | (0.507, 0.671) | (0.348, 0.589) | (0.159, 0.617) | (-0.066, 0.788) | (-1.580, 2.055) | (-13.38, 9.803) | (-22.582, 10.37) |

**Table 3.** The estimates of period-specific HR of 100-days interval with protective factor ($log(HR_L) < 0$)

| Treatment effect | Protective factor effect | True marginal effect | Estimated period-specific effect (day) | | | | | | | |
|---|---|---|---|---|---|---|---|---|---|---|
| $log(HR_E)$ | $log(HR_L)$ | $log(HR_M)$ | $log(\widehat{HR}_{unadj})$ (95% CI) | | | | | | | |
| | | | 1 - 100 | 101 - 200 | 201 - 300 | 301 - 400 | 401 - 500 | 501 - 600 | 601 - 700 | 701-800 |
| -0.9 | -0.4 | -0.823 | -0.91 | -0.875 | -0.849 | -0.81 | -0.787 | -0.744 | -0.7 | -0.727 |
| | | | (-1.377, -0.443) | (-1.193, -0.557) | (-1.179, -0.519) | (-1.202, -0.417) | (-1.259, -0.316) | (-1.348, -0.14) | (-1.531, 0.133) | (-2.806, 1.353) |
| | -0.2 | -0.865 | -0.901 | -0.879 | -0.868 | -0.848 | -0.796 | -0.800 | -0.795 | -0.697 |
| | | | (-1.218, -0.583) | (-1.111, -0.648) | (-1.109, -0.628) | (-1.129, -0.568) | (-1.153, -0.440) | (-1.289, -0.311) | (-1.438, -0.079) | (-1.68, 0.286) |
| -0.5 | -0.4 | -0.453 | -0.507 | -0.477 | -0.463 | -0.448 | -0.435 | -0.406 | -0.372 | -0.371 |
| | | | (-0.920, -0.095) | (-0.753, -0.201) | (-0.766, -0.16) | (-0.803, -0.093) | (-0.868, -0.002) | (-0.971, 0.160) | (-1.152, 0.409) | (-1.412, 0.67) |
| | -0.2 | -0.479 | -0.498 | -0.484 | -0.479 | -0.468 | -0.437 | -0.432 | -0.411 | -0.388 |
| | | | (-0.771, -0.224) | (-0.688, -0.28) | (-0.692, -0.266) | (-0.729, -0.207) | (-0.778, -0.096) | (-0.873, 0.008) | (-1.067, 0.245) | (-1.418, 0.643) |
| 0 | -0.4 | 0 | -0.002 | 0.006 | -0.001 | -0.001 | 0.015 | -0.011 | -0.008 | 0.049 |
| | | | (-0.349, 0.349) | (-0.240, 0.253) | (-0.273, 0.267) | (-0.299, 0.294) | (-0.395, 0.377) | (-0.525, 0.52) | (-0.700, 0.709) | (-2.065, 1.996) |
| | -0.2 | 0 | -0.002 | 0.006 | -0.001 | -0.001 | 0.015 | -0.011 | -0.008 | 0.049 |
| | | | (-0.231, 0.227) | (-0.182, 0.193) | (-0.190, 0.187) | (-0.230, 0.228) | (-0.295, 0.325) | (-0.451, 0.428) | (-0.690, 0.674) | (-1.046, 1.144) |
| 0.5 | -0.4 | 0.442 | 0.495 | 0.481 | 0.443 | 0.416 | 0.379 | 0.37 | 0.327 | 0.319 |
| | | | (0.183, 0.807) | (0.253, 0.709) | (0.197, 0.69) | (0.128, 0.705) | (0.010, 0.748) | (-0.144, 0.884) | (-0.436, 1.090) | (-0.673, 1.311) |
| | -0.2 | 0.474 | 0.495 | 0.487 | 0.472 | 0.436 | 0.423 | 0.387 | 0.265 | -1.402 |
| | | | (0.186, 0.805) | (0.252, 0.722) | (0.239, 0.704) | (0.101, 0.77) | (-0.021, 0.866) | (-0.314, 1.088) | (-2.467, 2.997) | (-14.998, 12.193) |
| 0.9 | -0.4 | 0.786 | 0.887 | 0.853 | 0.785 | 0.729 | 0.669 | 0.619 | 0.594 | 0.553 |
| | | | (0.597, 1.178) | (0.638, 1.067) | (0.548, 1.022) | (0.447, 1.01) | (0.303, 1.035) | (0.098, 1.14) | (-0.131, 1.320) | (-0.427, 1.533) |
| | -0.2 | 0.849 | 0.891 | 0.866 | 0.822 | 0.768 | 0.747 | 0.669 | 0.631 | -2.169 |
| | | | (0.699, 1.083) | (0.714, 1.018) | (0.659, 0.985) | (0.534, 1.003) | (0.422, 1.073) | (0.162, 1.177) | (-0.289, 1.552) | (-15.578, 11.24) |

**Figures**

**Figure 1.** The mean values of the harmful factor ($HR_L = 1.49$) for the treated and control groups over follow-up.

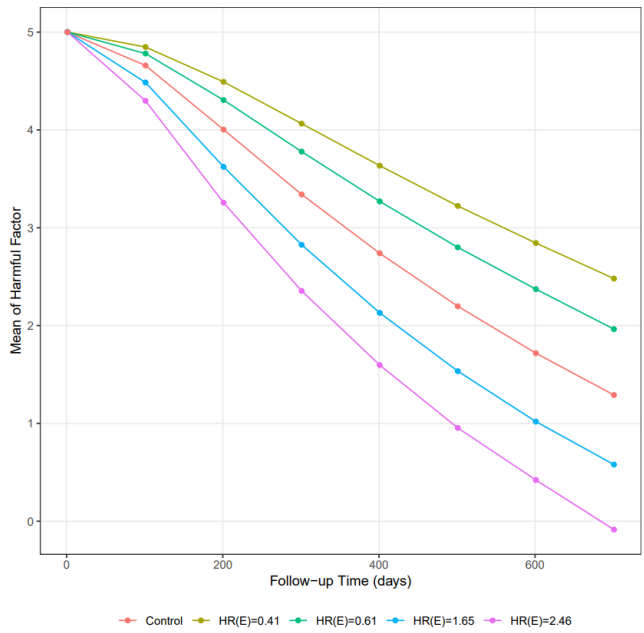

**Figure 2.** The mean values of the protective factor ($HR_L = 0.67$) for the treated and control groups over follow-up.

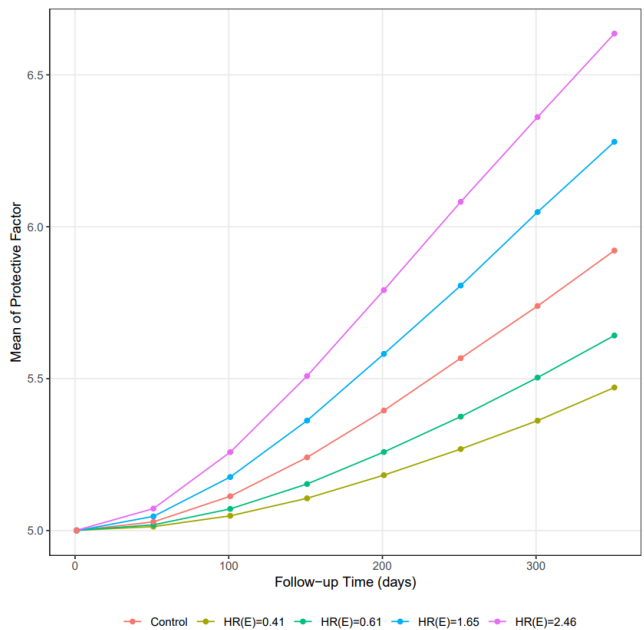